# MULTI-HOP BANDWIDTH MANAGEMENT PROTOCOL FOR MOBILE AD HOC NETWORKS

Binod Kumar Pattanayak, Manoj Kumar Mishra,
Alok Kumar Jagadev, Manojranjan Nayak

Siksha O Anusandhan University, Jagamohan Nagar, Khandagiri, Bhubaneswar
Orissa, India, Postal Zip : 751030, Tel:+919861291002, Fax:+916742351880.
E-mail: bkp_iter@yahoo.co.in
E-mail: mkmishra_iter@yahoo.com
E-mail: a_jagadev@yahoo.co.in
E-mail: president@soauniversity.ac.in

## ABSTRACT

*An admission control scheme should play the role of a coordinator for flows in a data communication network, to provide the guarantees as the medium is shared. The nodes of a wired network can monitor the medium to know the available bandwidth at any point of time. But, in wireless ad hoc networks, a node must consume the bandwidth of neighboring nodes, during a communication. Hence, the consumption of bandwidth by a flow and the availability of resources to any wireless node strictly depend upon the neighboring nodes within its transmission range. We present a scalable and efficient admission control scheme, Multi-hop Bandwidth Management Protocol (MBMP), to support the QoS requirements in multi-hop ad hoc networks. We simulate several options to design MBMP and compare the performances of these options through mathematical analysis and simulation results, and compare its effectiveness with the existing admission control schemes through extensive simulations.*

## KEYWORDS

*Ad hoc network, admission control, bandwidth, QoS, MBMP.*

## 1. INTRODUCTION :

As routing in ad hoc networks is organized with the cooperation among the nodes of the network, the nodes must cooperate with each other to provide QoS support too. Such cooperation includes implementation of certain policies at the endpoints of flows and implementation of an admission control policy along the routes, to make it possible for new flows consuming limited resources from the network not to disrupt the resources of existing flows. The principal goal of our research is to provide an efficient scalable admission control protocol for wireless ad hoc networks, so as to maintain the QoS support for end-to-end connections, with well-policed flows. The requirement of any QoS support is to provide applications with guarantees at the level of bandwidth, delay and jitter. Many solutions have been proposed for QoS support in wired networks and multichannel wireless networks like TDMA or CDMA. But, the physical characteristics of single-channel wireless networks like IEEE 802.11 networks [1] do not adhere to such solutions. The Central Problem around QoS support in single-channel wireless networks represents the shared nature of the wireless medium. In multi-hop ad hoc networks, nodes cannot communicate directly, and may still lead to some contention with each other for the same shared resource. Such a contention, which we name as c-neighborhood in this paper, can affect resource allocation at individual nodes of the network in two different ways : (i) decision of resource allocation at an individual node needs information about the allocated bandwidth along the entire route of communication and the bandwidth allocated to nodes, existing beyond its transmission range; (ii) contention along a route may involve multiple nodes. Our admission control scheme is devised taking into account such characteristics of single-channel wireless networks.

We present a Multi-hop Bandwidth Management Protocol (MBMP), which deals with admission control of flows in a single-channel ad hoc network, taking into consideration the available local resources at a node and the effects of admitting a new flow at the neighboring nodes. We concentrate on single channel ad hoc networks, based on single-channel MAC layers such as IEEE 802.11. In this paper, the description and analysis of MBMP are carried out with the implementation of IEEE 802.11 protocol, although it could be combined with other single-channel MAC layer protocols such as IEEE 802.11e [2] and SEEDEX [3].

The following sections are organized as follows. Section 2 specifies the motivation to the current work. The characteristics of wireless communication related to admission control in ad hoc networks are discussed in section 3. Section 4 comprises the challenges and corresponding solutions to implementation of admission control in ad hoc networks. Detail design of our proposed protocol, MBMP, is covered in section 5. An analysis of overhead pertained to admission control in c-neighborhoods is carried out in section 6. Comparison of simulation results of MBMP is made with respect to SWAN[4] and DSR[5] in section 7. Section 8 includes acknowledgement. In section 9, we conclude the paper and propose the future work.

## 2. MOTIVATION:

Bandwidth management for multi-hop communications across MANETs imposes a great concern, especially for real time traffic and multimedia applications. A wide range of solutions have been proposed in the literature for bandwidth management of single-hop communications in ad hoc networks, and not many for that in multi-hop communications. It is our endeavor to devise a protocol, which could successfully deal with the issues related to bandwidth management and admission control for multi-hop flows in mobile ad hoc networks.

## 3. CHARACTERISTICS OF WIRELESS COMMUNICATION:

Proper allocation of available communication resources can enhance efficient communication over a shared wireless medium. The problem of QoS support becomes even more complicated in an ad hoc network with dynamically moving mobile nodes. This section focuses on the challenges and current research on this field.

### 3.1. Wireless Channels:

The wireless networks principally differ from wired networks in openness of wireless networks. In wired networks, at any point of time, only authorized devices can use the communication channel, whereas wireless links are inherently shared. Any wireless node can simply send data and contend from the wireless channel. At the same time, there is no isolation from the other sources, which might be using the wireless channel with entirely different infrastructure (e.g. IEEE 802.11 & IEEE 802.15) or simply causing noise (e.g. microwave oven).The uniqueness of the structure of wireless channels leads to the following two challenges:

a)    Perception of Available Bandwidth: A node in a shared medium wireless network, starting to transmit a flow, consumes bandwidth at its c-neighbors. As each individual node has a different perception of the wireless network, it cannot determine on its own, if its c-neighbors have enough available bandwidth in reserve. In addition, a node cannot obtain the bandwidth information of a c-neighbor, which it cannot directly communicate with, and which lies within the carrier-sensing range of the node, but beyond its transmission range.

b)    Perception of Flow Bandwidth Consumption: Since multiple wireless nodes along a route may contend for bandwidth at a single location and do not know each other's bandwidth information, it is difficult for a node along the route of a flow, to determine how much bandwidth the flow would consume at its c-neighbors.

In the next section, we discuss the possible solutions to the above challenges.

International Journal of Managing Information Technology (IJMIT), Vol.2, No.4, November 2010### 3.2. Mobility:

Due to the mobility of nodes in ad hoc networks it becomes difficult to maintain a strict QoS. In addition, as the communicating nodes move into each other's transmission range, available bandwidth in the network decreases accordingly. Thus, QoS requirements in ad hoc networks should be relaxed to allow a better-than-best-effort service [6]. MBMP provides a QoS commitment, where a node does not necessarily break QoS requirements intentionally by admitting too many flows. But, when the commitment is broken due to mobility of a node, it sends a message to the source node regarding the changes in route. In such a case, the source may either search for a new route or reduce QoS requirements for the broken or degraded route. Due to mobility of nodes in ad hoc networks, the information gathered about the network by a node has a limited lifetime. Hence the information should be collected only when it is required. Our MBMP supports an on-demand admission control, where the message overhead is tied to the presence of flows.

### 3.3. Related Work:

The challenges, faced in wireless mobile ad hoc networks, refer to different aspects of restrictions over QoS requirements. Several approaches incorporating TDMA-based protocols have been proposed to support QoS requirements in wireless ad hoc networks [7, 8, 9, 10, 11], those are oriented around an effective synchronization among all nodes in the network and implement a slot allocation algorithm, which is vulnerable to mobility in the network.

Single Channel MAC layer scheduling algorithms have been suggested for resource allocations in wireless ad hoc networks [3, 12, and 13], which implement a single channel, shared by all the nodes of the network, where QoS requirements are maintained by a coordination of transmission schedules of packets among the nodes. These approaches are more flexible even in presence of mobility, since they support localized decisions at packet level and deal with fair resource allocation at the level of individual nodes.

A number of admission control schemes for mobile ad hoc networks have been proposed[4, 6, 14, 15, 16, 17, 18, 19]. However, some proposed solutions like INSIGNIA [14], MMWN[16] and a connectionless routing architecture [17] deal with only high level issues and do not consider the available bandwidth at the c-neighbors. Adaptive QoS Routing Algorithm (ADQR) [29] uses signal strength to predict route failures and obtains estimated bandwidth from lower layers. Optimized Link State Routing (OLSR) [30]protocol always finds a path with larger available bandwidth. The solutions provided by SWAN[4], VMAC [6], [15], [18], [19] do not pay any attention to the resources of the c-neighbors of a node during admission control, rather focus on only the local resources. However a node may need to consume resources of its c-neighbors through contention. In our research work, we have tried to overcome the above problems by allowing our MBMP to consider both local resources as well as resources at c-neighbors, while making decisions on admission control.

## 4. MULTI-HOP BANDWIDTH MANAGEMENT:

The goal of admission control is to determine if the available resources can suffice to the requirements of a new flow to be admitted, not disrupting the bandwidth levels of the existing flows. To achieve this goal in ad hoc networks, we address two challenges, discussed in section 3.1, i.e. perception of available bandwidth and perception of bandwidth consumption of a flow.

### 4.1. Perception of Available Bandwidth:

The first challenge to MBMP is the evaluation of available bandwidth in the network where bandwidth requirements of all the flows taken together do not exceed the available resources in the network. We introduce two parameters: c-neighborhood available bandwidth and local available bandwidth. The c-neighborhood available bandwidth is defined as the maximum amount of bandwidth that a node can use without disrupting the reserved bandwidth of any existing flows in its carrier-sensing range (c-neighborhood). Local available bandwidth is defined as the amount of unused bandwidth at a node at any point of time. Hence, to admit a flow successfully, a node requires sufficient local and c-neighborhood available bandwidth.

We demonstrate a simulation using NS2 [20] with six mobile hosts with an orientation as shown in Fig. 1. It uses MAC layer protocol IEEE 802.11, with a radio transmission range of 250m and a carrier sensing range of 550m. The wireless channel has a bandwidth of 2 Mbps. Nodes C and E are c-neighbors of each other. Node A is node C's neighbor and lies beyond the carrier – sensing range (c-neighbor) of node E. Three CBR flows (Flow 1, Flow 2, and Flow 3) are established between node pairs A-B, C-D and E-F respectively, with a transmission rate of 133 packets per second and a packet size of 512 bytes. Each of the flows requires a channel bandwidth of around 930 kbps, taking into account the overhead of RTS-CTS-DATA-ACK [21] handshakes and collisions (Fig.2) at the MAC layer. At time t1 seconds, node A initiates Flow1 to node B, at time t2=40 seconds, node C initiates Flow 2 to node D, and at time t3 = 80 seconds, node E initiates Flow 3 to node F. Fig. 3 demonstrates the changes in local available bandwidth at each source node as the three flows are initiated successively. Fig.4 and Fig.5 depict the throughput and delay incurred by each of the three flows respectively.

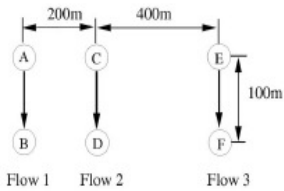 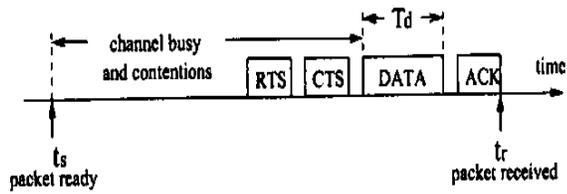

Fig. 1: Simulation topology  Fig.2: RTS-CTS-DATA-ACK handshake

Fig. 3: Changes of local available bandwidth

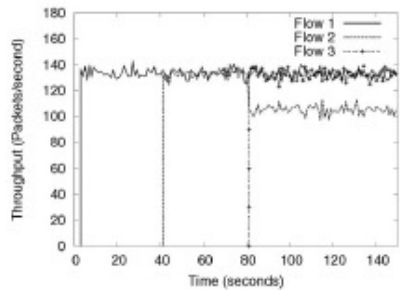 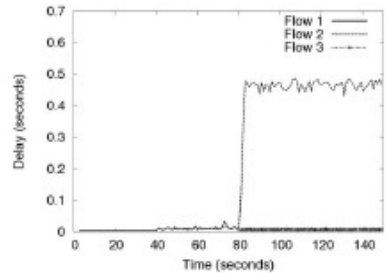

Fig. 4: Throughput of Flows 1, 2, and 3.  Fig. 5: Delay of Flows 1, 2, and 3

As shown in Fig.3, after Flow 2 starts, node E has a local available bandwidth of 1.07 Mbps, which is not consumed by contention from Flow 2. This much of local available bandwidth is sufficient to admit Flow 3. Consequently Flow 3 can very much achieve its desired throughput and delay (Fig. 4 and Fig. 5). In the earlier approaches to admission control [6, 4, 15, 18, 19], as no consideration is taken for c-neighborhood available bandwidth, Flow 3 would be successfully admitted. However, since node E lies in the c-neighborhood of node C, node E can be able to consume local available bandwidth of node C. In this scenario, when Flow 3 starts, the contention from it causes the throughput of Flow 2 to decrease by 20 percent and increases the delay of Flow 2 convincingly (Fig. 4). The decrease in QoS requirements of flow 2 is resulted as a consequence of node C having only 0.14 Mbps of local available bandwidth before Flow 3 starts, which is much less than the bandwidth consumed by Flow 3. Otherwise, node E does not possess enough c-neighborhood available bandwidth to admit Flow 3. Hence, to admit a flow, only the local available bandwidth can not suffice to the requirements, rather the node should possess enough c-neighborhood available bandwidth too.

### 4.1.1. Calculation of Local Available Bandwidth:

The local available bandwidth to a node can be defined as the unconsumed bandwidth at the node, which is to be determined by the node, by passively monitoring the activities across the network. In the current paper, we take into account the history of idle channel time, which can contribute to calculation of local available bandwidth at the node. However, the idle channel time caused by the back-off algorithm of IEEE 802.11 and collisions in the network cannot fulfill the purpose. But, as suggested in [22], the amount of idle channel time and collision time in IEEE 802.11 is much negligible as compared to packet transmission time. Hence, considering the idle channel time for calculation of local available bandwidth can be a simple approximation of it. In [18] and [19], the authors suggest predicting the local available resources using the reciprocal of the current transmission delay, which consequently determines the local achievable bandwidth. But, local achievable bandwidth refers to the maximum amount of bandwidth that a flow can be able to achieve competing with the existing flows, which may contradict to the purpose of admission control, as admission control should not consequently disrupt the throughput of the existing flows. Hence, in our scheme, MBMP purposefully uses idle channel time for estimation of local available bandwidth.

More or less, the channel at a node can be perceived to be either busy or idle. The channel can be perceived to be busy at any point of time, if the node is not in any one of three states: (i) the node is transmitting or receiving a packet; ii) the node is receiving a RTS or CTS message [21] from another node; (iii) the node is sensing a busy carrier with a signal strength much larger than a predefined threshold, called the carrier-sensing threshold, but the node is unable to interpret the contents of the message. After monitoring the amount of idle channel time, $T_{idle}$, during every period of time, $T_p$, the local available bandwidth, $B_{local}$, for a given node can be calculated using a weighted average as in equation (1):

$$B_{local} = \alpha B'_{local} + (1-\alpha) \frac{T_{idle}}{T_p} B_{Channel} \quad \ldots\ldots\ldots\ldots\ldots\ldots\ldots\ldots (1)$$

Where $B'_{local}$ is the local available bandwidth of the node in the preceding period, which is initially zero, $B_{channel}$ is the channel capacity in bits per second and the weight $\alpha \in [0, 1]$.

In MAC layer protocols like IEEE 802.11e[2] which support priority based flow scheduling, the estimation of local available bandwidth requires estimation of the amount of bandwidth consumed by existing low priority flows, since to admit high priority flows, low priority flows have to be deprived of their allotted bandwidth, if needed. In [23], a method is presented to calculate local available bandwidth for each priority level, which can be easily incorporated with MBMP.

### 4.1.2. Calculation of c-Neighborhood Available Bandwidth:

To get an estimate of c-neighborhood available bandwidth, a node can implement any one of the two approaches, namely, active approach and passive approach. In active approaches, c-neighbors actively exchange the information about bandwidth between each other. In passive approaches, a node has to monitor the channel passively, to obtain the c-neighborhood available bandwidth. In the current paper, we bring forward two active approaches and one passive approach for estimation of bandwidth information at c-neighbors of a node.

The MBMP-Multi-hop approach, an active approach, enables a node to broadcast queries with a limited hop count, to contact all its c-neighbors. However, limited hop count represents the bottleneck of this approach, as it could be difficult to reach all the c-neighbors with a small hop count in some specific topologies. In Fig.6, if node A sends queries limited to 2 hops, it cannot reach nodes E and G, since they lie beyond the limited hop count, even though E & G are in the carrier-sensing range of node A. If queries are sent with a limited hop count of 3, node H can be virtually included, although it lies beyond the carrier-sensing range of node A. In addition to this, node B cannot be reached by node A, irrespective of how much the hop count is limited to, although node B is a c-neighbor of node A. The cost of a query in this approach is directly related to the number of hops required to reach the c-neighbors. In the beginning, the querying node sends a message, and consequently, all nodes, that are one hop away from the querying node, receive the message and broadcast the message further. The carrier-sensing range in IEEE 802.11 is twice the transmission range, and thus, in our simulations using IEEE 802.11, MBMP-multi-hop approach implements a hop count of 2 as its transmission range.

The MBMP–Power approach uses the benefits of power control capabilities of modern wireless technology. In this approach, a sender can take the advantage of using a larger transmission power level for its queries than the power level required for normal data transmission. This additional power level is obtained using additional hardware. In this approach, the queries of sender can reach all of its c-neighbors. It should be noted that, this technique is required only for sending bandwidth information, which occurs less frequently as compared to frequency of data transmission. Transmission of data messages occurs at normal transmission power level, since data transmission with enhanced power level may lead to reduction of network capacity. [24].

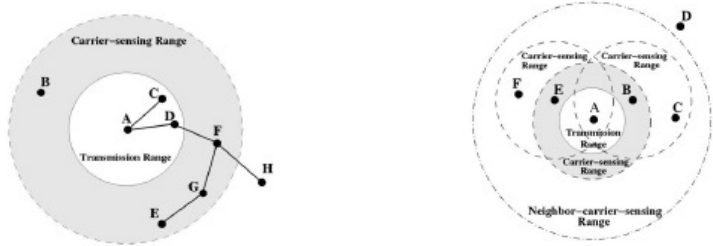

**Fig. 6: Multi-hop approach    Fig.7: Neighbor-carrier sensing approach**

The MBMP-CS approach, a passive approach, does not implement any query message to c-neighbors. With normal IEEE 802.11 operations, a node can passively monitor the medium using a Neighbor-carrier sensing Threshold, which is set to much lower than the carrier–sensing threshold. The sensing range using this threshold, called Neighbor-carrier – sensing range, covers the carrier sensing ranges of all of the sensing nodes c-neighbors as shown in Fig.6. When a node senses the carrier strength to be smaller than the Neighbor-carrier-sensing Threshold, then there is no communication activity in its c-neighborhood and all c-neighbors of the node sense idle channels. With the amount of time that the channel is in idle neighbor state, $T_{idle}^{neighbor}$ for every period of time, $T_p$, the c-neighborhood available bandwidth, $B_{neighbor}$, can be derived using the following formula:

$$B_{neighbor} \approx \propto B'_{neighbor} + (1-\propto)\frac{T_{idle}^{neighbor}}{T_p} B_{Channel} \quad \dots\dots\dots\dots\dots\dots\dots\dots\dots\dots\dots\dots\dots (2)$$

Where $B'_{neighbor}$ is the c-neighborhood available bandwidth in the preceding period of time, initially approximated to zero.

International Journal of Managing Information Technology (IJMIT), Vol.2, No.4, November 2010The variant of MBMP with this approach is known as MBMP-CS approach. As depicted in Fig.7, estimation of c-neighborhood available bandwidth in MBMP-CS approach is more conservative, even though it has the lowest message overhead as compared to MBMP-multi-hop and MBMP-power approaches. As shown in Fig.7, nodes E and B are in the carrier-sensing range of node A. Node C is in the carrier-sensing range of node B, whereas node F lies in carrier-sensing range of node E. Nodes C and F are beyond the carrier-sensing range of node A and within Neighbor-carrier-sensing range of node A. Let the channel capacity be 2 Mbps and nodes C and F are transmitting at a rate of 1 Mbps each. As the local available bandwidth at nodes E and B are 1 Mbps each, the c-neighborhood available bandwidth at node A is 1 Mbps too. However, when either node C or node F transmits, node A can sense the channel not to be in idle neighbor state. For this reason, as long as the transmissions of nodes C & F do not overlap, the estimated c-neighborhood available bandwidth of node A, using Neighbor-carrier-sensing Threshold, will be less than 1 Mbps, since by monitoring the channel, node A could not know that node C lies beyond the carrier-sensing range of node E and does not consume the bandwidth of node E. Hence, node A can only presume that any transmission activity in its Neighbor-carrier-sensing range can consume bandwidth of all of its c-neighbors.

Evaluations of all three approaches of MBMP including their accuracy and message overhead are discussed in section 6.

### 4.2. Bandwidth Consumption:

The second challenge to MBMP represents the estimation of bandwidth to be consumed by a new flow so as to decide whether the available bandwidth can suffice to the requirements of the new flow. In the beginning, the transmission rate of the application should be mapped into the respective channel bandwidth requirement. In course of the mapping, the protocol overhead of the MAC layer as well as the networking layer must be taken into consideration. As an instance, IEEE 802.11 protocol, used at the MAC layer for each application data packet, must implement RTS-CTS-DATA-ACK handshake [21]. Hence, the transmission time for each data packet, $T_{data}$, can be calculated as :

$$T_{data} = T_{difs} + T_{rts} + T_{cts} + \frac{L+H}{B_{Channel}} + T_{ack} + 3\, T_{sifs} \quad \dots\dots\dots\dots\dots\dots\dots\dots\dots\dots\dots\dots\dots\dots (3)$$

Where L is the size of data packets, H is the length of IP header and MAC packet header taken together, $T_{rts}$, $T_{cts}$ and $T_{ack}$ are the time for transmission of RTS, CTS and ACK packets, respectively. $T_{sifs}$ and $T_{difs}$ represent the inter frame spaces for short inter frame spacing (SIFS) and distribution coordination function spacing (DIFS) respectively. For R packets with an average packet size of L generated by the application per second, the respective channel bandwidth requirement, W, of the flow can be calculated as :

$$W = R \times T_{data} \times B_{channel} \quad \dots\dots\dots\dots\dots\dots\dots\dots\dots\dots\dots\dots\dots\dots\dots\dots\dots\dots (4)$$

In addition, multiple nodes along the route of a flow may contend for bandwidth at a single location, and as a result, each of these nodes consumes an amount of bandwidth equal to W at this location. The number of such nodes contributing to a contention is known as contention count of the route and is denoted as $N_{ct}$. Thus, amount of bandwidth consumed by the flow at this location, $B_c$, can be calculated as :

$$B_c = N_{ct} \times W \quad \dots\dots\dots\dots\dots\dots\dots\dots\dots\dots\dots\dots\dots\dots\dots\dots\dots\dots\dots\dots\dots\dots\dots (5)$$

As shown in Fig. 8, Flow 1 operates along route A→B→ C→D→E. Nodes A, B, C, D and E contend for bandwidth at node F, since they lie within carrier-sensing range of node F, and hence, the contention count at node F is 5. If Flow 1 requires a bandwidth of 2 kbps, then it consumes a minimum of 10 kbps at node F.

Assuming a node Q being c-neighbor of n transmitting nodes along the route of a flow, which requires W Kpbs of channel bandwidth, if Q lies along the route of the flow, then the bandwidth consumed at node Q is (n+1)W kbps and the contention count at node S is (n+1), and if Q does not lie along the route of the flow, then bandwidth consumed at node Q is nW kbps and the contention count at node Q is n. Node Q needs to identify its c-neighbors, called c-neighbor set ($S_{c-nb}$) to calculate its contention count. For a given c-neighbor set of node Q, $S_{c-nb}$, and for a given route of a flow, the contention at Q, $N_{ct}$, can be determined as the number of common nodes participating in transmission of the flow and c-neighbor set of node Q, which can be expressed as

$$N_{ct}(Q) = \begin{cases} |(Route - Dest) \cap S_{c-nb}|, & if\ Q \notin Route \\ |(Route - Dest) \cap S_{c-nb}| + 1, & if\ Q \in Route \end{cases} \quad \ldots\ldots\ldots(6)$$

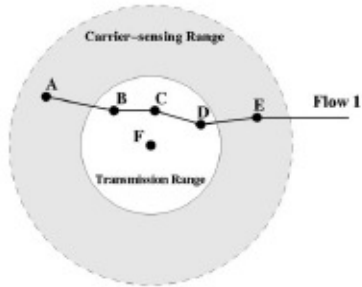

TABLE 1
Configurations of CBR Sources

| Flow No. | Rate (Pkts/s) | Packet Size (Bytes) | Starting Time (s) |
|---|---|---|---|
| 1 | 13.5 | 112 | 10 |
| 2 | 42.65 | 381 | 20 |
| 3 | 35.55 | 311 | 30 |
| 4 | 16.99 | 481 | 60 |
| 5 | 37.69 | 519 | 80 |
| 6 | 18.69 | 855 | 90 |
| 7 | 44.04 | 317 | 100 |
| 8 | 46.20 | 786 | 110 |
| 9 | 14.92 | 402 | 140 |

**Fig.8: Neighbor-carrier sensing approach**

Where the destination does not contend for the channel, since it only receives traffic passively and does not need to send. The c-neighbor set can be determined using both active and passive methods. In active method, each node broadcasts a hello message periodically to its one-hop neighbors. This hello message incorporates the identities of the initiator as well as its k-hop neighbors (k=1, 2, 3….) with hop counts obtained via broadcast messages from other nodes. After receiving a hello message, a receiver node fetches the identities of the initiator and the k-hop nodes in its c-neighbor set. Consequently, every node learns its (k+1)-hop c-neighbors' identities. For a given route of a flow, a node knows about how many nodes along the route of the flow belong to its k-hop c-neighborhood. In passive method, the c-neighbors can be learned by passively monitoring the routes and the initiator's information contained in data and control messages, which effectively reduces the communication overhead. It becomes relevant, as MBMP uses source routing from the reasons specified in section 5.1. For example, in Fig. 8, node F learns about node D as its one-hop neighbor, when node F can listen to a message sent by node D to node E along the source route A→B→C→D→E. Node F can also infer that nodes E and C are at most two hops away and, node B is at most three hops away and node A is at most four hops away from the source route information. At the same time, if node F can listen to a message from node B along the same route, then node F will require to update its distance to node B as one hop and that to node A is two hops. Thus, node F gradually learns the accurate distance to its c-neighbors by monitoring the traffic around itself. Knowing the identities of nodes within k hops in its c-neighbor set, node F can be able to determine the contention count of a flow. A major concern around this method is that the c-neighbor set may not be complete by the time when bandwidth consumption of a flow must be calculated to perform admission control. In section 5, it is demonstrated with an example that the c-neighbor set built through the passive method is complete enough by the time when an accurate estimation of contention count of a flow is needed.

## 5. MBMP DESIGN:

MBMP (Multi-hop Bandwidth Management Protocol) incorporates single channel MAC layer protocols like IEEE 802.11 [1], IEEE 802.11e [2] and SEEDEX [3] and performs admission control with bandwidth aware routing. Routing in MBMP involves low message overhead in the presence of mobility and implements local approach for estimation of bandwidth consumption. MBMP constitutes four components: route discovery, admission control, and building c-neighbor sets.

### 5.1. Route Discovery:

The intention behind route discovery is to determine a route from sender to receiver that possesses sufficient resources for initiation of a flow. MBMP implements on-demand route discovery with source routing, similar to DSR [5]. We use the effectiveness of source routing-based approach, as it allows MBMP to specify directly the route, along which the packets of the flow move, and which is determined by admission control and has enough reserved bandwidth for initiation of the flow. Routing protocols like DSDV [25], AODV [26] and TORA [27] do not necessarily implement source routing and may route the packets of the flow to some other route if sufficient resources are not available.

MBMP implements partial admission control in the process of route discovery to reduce the message overhead, so as to eliminate routes with insufficient for the flow available bandwidth. In this approach, prior to sending data, a source node broadcasts a route request message to its neighbors. This route request message contains the bandwidth requirement of the connection, computed as in equation (4), the address of the initiating node, address of the destination node, and a record of sequence of hops of the route, along which route request message is sent via the ad hoc network. This sequence of hops, known as partial route (PRoute) is used to determine the lower bound of contention count along the entire route and to eliminate the circular routes. Every node that receives the route request message implements partial admission control to determine if there is enough available network bandwidth to admit the flow along the partial route. In case partial admission control does not succeed or the partial route contains loops, the route request is rejected. Otherwise, the node appends its own address to the partial route and rebroadcasts the route request message.

If the route request message reaches the intended destination node, the partial route in the route request becomes a full route. Following it, the destination node reverses the full route and sends a route reply message back to the initiating source node along the same route. In case multiple route request messages carrying different routes arrive at the destination node, the destination node sends a route reply message only along one of the routes, selected with reference to some selection criterion like shortest route or first route request etc. However, the other routes are cached as backup routes in case the first route reply message does not reach the initiating node due to link failure or admission failure. At each node along the route of the route reply message, full admission control is performed. When admission control succeeds at a node, a soft reservation of bandwidth is set up and the route reply message is forwarded to the next hop along the route. Otherwise, an admission failure message is returned to destination node. When the nodes along the path of the admission failure message to destination receive this message, the soft reservations of bandwidth at these nodes implemented earlier are dropped explicitly. After receiving the admission failure message, the destination node selects another cached route and sends a route reply message along this route. If the route reply message successfully reaches the source node, enough end-to-end bandwidth along the route must have been reserved and the communication can start.

### 5.2. MBMP Admission Control Algorithm:

In the process of route discovery, multiple possible routes can be determined by the source node to reach a destination. Consequently, admission control must be used to identify a route which can admit the new flow. At each node along the route, decision on admission control can be carried out on the basis of expected bandwidth consumption of the flow as well as the available bandwidth at the node and its c-neighbors. MBMP implements admission control in two phases of route discovery. (1) Partial admission control is performed during route request, when each node along the route receives route request message; (2) Full admission control is performed during route reply, when each node along the route receives a route reply message. This separation of admission control in two phases is necessary, since during route request phase, the entire route to the destination is unknown yet. The expected bandwidth consumption of the flow calculated on the basis of partial route carried in the route request message may be smaller than the actual bandwidth consumption of the final route, since the contention count of the final route cannot be calculated in this phase. Hence, admission control in this phase cannot be complete due to inaccurate estimated bandwidth consumption of the flow, and thus, is called partial admission control. Partial admission control is used as a first pass to weed out routes and reduce message overhead by avoiding multiple route requests around hot spots in the network, since it may be over-optimistic in admitting flows. In section 7, the effectiveness of partial admission control is clearly demonstrated via simulations. Full admission control provides an accurate admission control, since during route reply phase, the full route to destination is known.

### 5.2.1. Partial Admission Control:

In MBMP, after receiving a route request message, a node implements partial admission control by comparing its available bandwidth with probably underestimated bandwidth consumption, calculated using partial route (equations (5)&(6)). However, types of available bandwidth used in partial admission control in three versions of MBMP, i.e. MBMP-multi-hop, MBMP-power and MBMP-CS, are different. In MBMP-multi-hop and MBMP-CS approaches, estimation of c-neighborhood available bandwidth involves identification of c-neighbors, which is an expensive operation. For this reason, querying c-neighbors should be avoided for the nodes, which do not lie along a viable route to destination. As route request messages are flooded across the entire network, to reduce overhead during a route request phase, only local available bandwidth for a node is calculated using equation (1) and compared with bandwidth consumption of the flow. If the local available bandwidth appears to be less than bandwidth consumption of the flow, then admission control is dropped. Otherwise, admission control is performed and route request message is forwarded to the next hop along the route to destination. In MBMP-CS approach, the bandwidth consumption of the flow is compared to both local available bandwidth and c-neighborhood available bandwidth taken together, since in this approach, estimation of c-neighborhood available bandwidth does not involve extra message overhead, and estimation of c-neighborhood available bandwidth is carried out with equation (2).

### 5.2.2. Full Admission Control:

Full admission control is performed by a node in route reply phase, when it receives a route reply message. At first, the local available bandwidth of the node is compared with calculated bandwidth of the flow at the node's location. Estimation of bandwidth consumption of the flow in this phase is accurate, since the route reply message in this phase incorporates the full route. If the local available bandwidth of the node is found to be higher than the bandwidth consumption of the flow, the c-neighborhood available bandwidth of the node is compared with the bandwidth consumption of the flow.

To achieve the above, the three approaches of MBMP use different approaches, as discussed in section 4.1.2. In MBMP-power and MBMP-multi-hop (both active) approaches, a node, after receiving route reply message broadcasts an admission request message (via multi-hop or enhanced power techniques), to its c-neighbors, which holds the full route of the flow. After receiving the admission request massage each of the c-neighbors of the node calculates the bandwidth consumption of the flow using equations (4), (5) and (6), and compares it with its local available bandwidth. If the local available bandwidth at the node is found to be smaller than the bandwidth consumption of the flow, then the node sends an admission rejection message back to the initiator and the admission control in c-neighborhood of the node fails. In case an admission rejection message is not received by the initiator in a stipulated period of time, it times out and the initiator assumes that the full admission control succeeds. The stipulated period of timeout is determined by the propagation delay, transmission time of admission rejection message and computation time. Some c-neighbors of the node may fail to receive the admission request broadcast message as a result of collisions. Since a node may belong to c-neighborhood of multiple nodes along a route, it is however, unlikely that a node fails to receive the admission request broadcast message. In Fig. 8, for example, node F lies in the c-neighborhood of nodes A,B,C,D, and E. Nodes A, B, C, D and E all broadcast admission request messages for Flow 1. Node F only needs to receive the admission request broadcast message from at least one of the above nodes for admission control to succeed.

Passive approach MBMP-CS does not incorporate any request / reject message during admission control of the flow and calculation of c-neighborhood available bandwidth is done using passive approach. A node, after receiving a route reply message, estimates its c-neighborhood available bandwidth using equation (2), compares it with bandwidth consumption of the flow, and accordingly decision can be taken regarding admission control. As elaborated in section 4.1.2, three versions of MBMP possess different message overhead and different levels of accuracy in full admission control.

### 5.3. Building c-Neighbor Set:

The accuracy of admission control strictly depends on the completeness and accuracy of c-neighbor set. Information about c-neighbors in MBMP can be gathered by monitoring control and data messages incorporated in c-neighbor set. In the absence of communication activity in a node's neighborhood for a long period of time, the c-neighbor set entries at this node may all time out. Consequently, if a new flow needs to go through this node, the node would not be able to give an accurate estimation of bandwidth consumption of the flow due to the absence of its c-neighbor set. Nevertheless, our following analysis shows that such a scenario happens rarely. When the route request messages are flooded across the entire network in the route request phase, and the route request message carrying the partial route it has traversed, reaches the node, the node can cache the last two hops of the partial route as its c-neighbor set and collect information about its c-neighbors. Again, as admission request messages in MBMP-multi-hop and MBMP-power approaches are sent to reach c-neighbors, the c-neighbor set can further be updated by caching the senders and forwarding nodes of these messages in the c-neighbor set. Finally, after receiving the route reply message, a node can add the last two forwarding nodes of this message to its c-neighbor set. Hence, till the moment a node needs to implement full admission control, its c-neighbor set must have been filled up optimally.

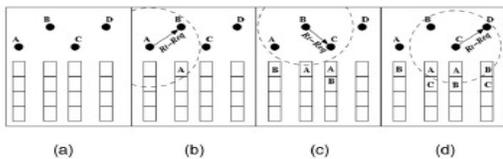
**Fig.9: Route request (MBMP-Multi-hop, MBMP-Power and MBMP-CS)**

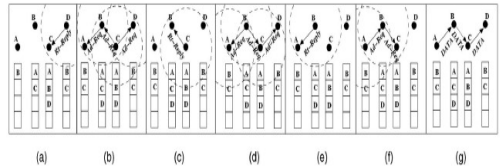
**Fig.10: Admission success of MBMP-Multi-hop**

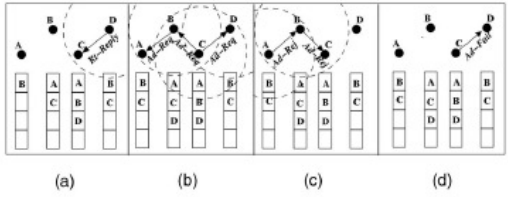
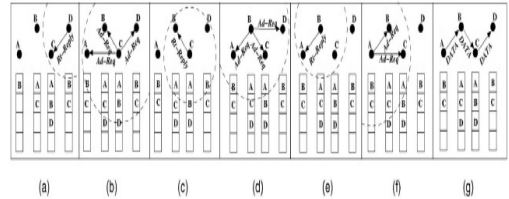

Fig.11: Admission failure of MBMP-Multi-hop

Fig.12: Admission success of MBMP-Power

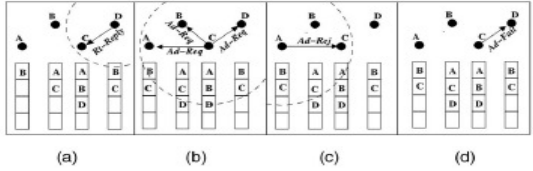
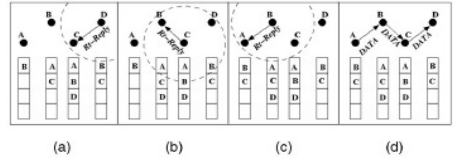

Fig.13: Admission failure of MBMP-Power

Fig.14: Admission success of MBMP-CS

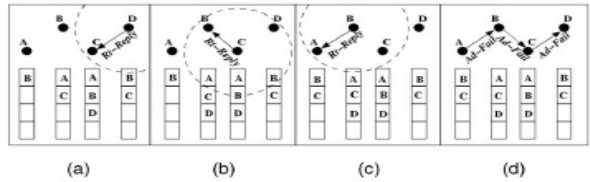
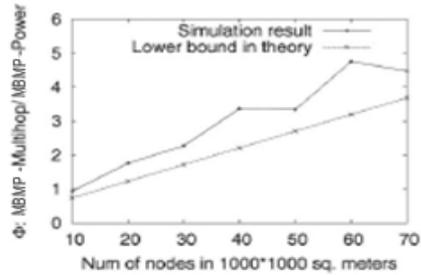

Fig.15: Admission failure of MBMP-CS

Fig.16: Overhead ratio of MBMP-Multi-hop versus MBMP-Power

## 5.4. An Example:

We illustrate the process of route discovery and admission control in three variants of MBMP (Fig.9,10,11,12,13, 14, 15). The tables under each node in these figures specify the c-neighbor set cached with the node. We presume in this example that the carrier-sensing range is twice the transmission range and hence the c-neighbor sets for nodes cache up to the 2-hop neighbors. To start with, the nodes do not receive any message and c-neighbor sets are empty (Fig. 9a).

In the beginning, node A initiates a connection to node D, for which it first maps the requirement of transmission rate of the connection to bandwidth requirement, W as per equation (4). Following it, node A initiates partial admission control. If the partial admission control succeeds, node A broadcasts a route request message with partial route {A} and a bandwidth requirement W (Fig. 9b). On receiving the route request message, node B adds node A to its c-neighbor set, updates the partial route to {A, B}, and performs partial admission control. With reference to equations (5) and (6), the bandwidth consumption at node B can be estimated to be:

$B_c = N_{ct}(B) \times W = [|(PRoute - Dest) \cap S_{c-nb}|+1] \times W = [|\{A,B\} \cap \{A\}| + 1] \times W = 2W$.

As only the partial route is known yet, $B_c$ is underestimated, as discussed in section 5.2. But, it does help to weed out all routes that begin with A→B, if it could be concluded that node B does not possess sufficient bandwidth. In MBMP-multi-hop and MBMP-power approaches (active), node B compares $B_c$ with its local available bandwidth as per equation (1). If its local available bandwidth is found to be greater than $B_c$, the partial admission control succeeds at node B and node B rebroadcasts the route request message with partial route {A, B} (Fig. 9c). As per MBMP-CS (Passive) approach, due to light weight estimation of c-neighborhood available bandwidth, node B compares its c-neighborhood available bandwidth with $B_c$ in the process of partial admission control, with reference to equation (2).

On receiving the route request broadcast message from B, node A adds node B to its c-neighbor set. As node A already exists in the partial route, it drops the route request message to avoid creating a circular route. On receiving the route request broadcast message from node B, node C adds both nodes A and B, available in the partial route, to its c-neighbor set and performs partial admission control similar to node B. If partial admission control succeeds at node C, it broadcasts route request message with partial route {A, B, C} (Fig. 9d). On receiving this route request broadcast message from node C, node B caches node C in its c-neighbor set and drops the route request message. On receiving the route request broadcast message from node C, the destination node D adds nodes B and C to its c-neighbor set and performs partial admission control. On successful partial admission control, node D reverses the route and sends a route reply message back to node C (Fig. 10a, 12a, 14a). At any point of time, in the process of partial admission control along the route A → B → C → D, if partial admission control does not succeed at a node, the partial admission control initiated by node A is dropped.

As per MBMP-multi-hop and MBMP-power approaches, when node C receives the route reply message, it performs full admission control by comparing its local available bandwidth with the bandwidth consumption of the flow, which is estimated as:

$$B_c = N_{ct}(B) \times W = [|(Route - Dest) \cap S_{c-nb}| + 1] \times W$$
$$= [|\{A,B,C,D\} - \{D\} \cap \{A,B,D\}| + 1] \times W = 3W.$$

If full admission control succeeds at node C, then it broadcasts an admission request message to all its c-neighbors via multi-hop or enhanced power (Fig.10b&12b). On receiving the admission request message from node C, node A adds node C to its c-neighbor set and calculates the bandwidth consumption of the flow as 3W since:

$$N_{ct}(A) = |(Route - Dest) \cap S_{c-nb}| + 1 = |(\{A,B,C,D\} - \{D\} \cap \{B,C\}| + 1 = 3$$

If node A's local available bandwidth appears to be insufficient to accommodate the flow, then admission request is dropped and a message, regarding the failure of admission of the flow, is sent to node C and then from node C to node D (Fig. 11a, 11b, 11c, 11d, 13a, 13b, 13c and 13d). If all the c-neighbors of node C possess enough local available bandwidth to admit the flow, then none of them sends admission rejection message. After node C times out, it sets up a soft reservation of the bandwidth and forwards a route reply message to node B. (Fig. 10C&12C). If full admission control succeeds at node B, then node B forwards the route reply message to node A. After a full admission control for the flow is accomplished at node A, the route (A→B→C→D) is supposed to have enough available bandwidth for the flow and transmission of data packets can resume (Fig. 10a,10b, 10c, 10d, 10e, 10f, 10g, 10h, 12a, 12b,12c, 12d, 10b, 12e, 12f, 12g, 12h).

It should be noted that in MBMP-CS, no admission request message is sent as depicted in Fig.14. On receiving a route reply message, a node performs full admission control by comparing the bandwidth consumption of the flow with the directly estimated c-neighborhood available bandwidth as per equation (2) and its local available bandwidth as well. In case there is enough c-neighborhood available bandwidth for the flow, then the route reply message is forwarded to next hop along the route until it reaches node A, as in Fig. 14. Along the route, if at any node, the c-neighborhood available bandwidth is found to be smaller than the bandwidth consumption of the flow, then an admission rejection message is sent back to the destination, as depicted in Fig. 15, and consequently, admission control for the flow is dropped.

## 6. ANALYTICAL EVALUATION AND SIMULATION RESULTS:

As MBMP-CS approach does not impose any additional message overhead for estimation of c-neighborhood available bandwidth, we compare the overhead of only MBMP-multi-hop and MBMP-power approaches. During the transmission of a control message using enhanced power level or multi-hop nodes, more interference is imposed to the network than in case of a normal message. To estimate the overhead in both MBMP-multi-hop and MBMP-power approaches, we use the total number of times that nodes receive admission request and admission rejection messages. In the process, we carry out an analytical evaluation followed by the simulations.

Let the transmission range of a node in a network with N nodes be R and the carrier sensing range be 2R. The density function $\rho(x,y)$ represents the number of nodes in a unit area centered at location $(x,y)$. Let $d_{xy}$ be a very small square region centered at location with area d. Hence the number of nodes in $d_{xy}$, say K, can be expressed as the product of area and the density, i.e.

$$K = \rho(x,y) \times d \quad \ldots\ldots\ldots\ldots\ldots\ldots (7)$$

Let the probability of each node in the network generating an admission request per unit time is q. Thus, the expected number of admission requests per unit time in region $d_{xy}$ can be calculated as

$$N_e = q \times \rho(x,y) \times d \quad \ldots\ldots\ldots\ldots\ldots (8)$$

Assuming that MBMP-multi-hop approach is used with 2 hops, it is more conservative in terms of overhead, but however, may not reach all nodes in carrier-sensing range. In this approach, an admission request message is heard by $\pi R^2 \rho(x,y)$ nodes and each of these nodes rebroadcasts the admission request message, and as a result the admission request message is received by a node $\pi R^2 \rho(x,y) + (\pi R^2 \rho(x,y))^2$ times. In MBMP-power approach, $4\pi R^2 \rho(x,y)$ nodes can hear the admission request message.

With respect to the above analysis, splitting the whole network into small square regions with area d, the expected ratio of overhead of the two approaches, $\theta$ can be estimated as:

$$\theta = \frac{\text{Message overhead of MBMP} - \text{multihop approach}}{\text{Message overhead of MBMP} - \text{power approach}}$$

$$= \lim_{d \to 0} \frac{\sum_{i=1}^{n} \left[ \pi R^2 \rho(x_i, y_i) + [\pi R^2 \rho(x_i, y_i)]^2 \right] N_{ei}}{\sum_{i=1}^{n} 4\pi R^2 \rho(x_i, y_i) N_{ei}} = \frac{1}{4} + \frac{\pi R^2}{4} \lim_{d \to 0} \frac{\sum_{i=1}^{n} \rho^2(x_i, y_i)}{\sum_{i=1}^{n} \rho(x_i, y_i)} \quad \ldots\ldots (9)$$

Where $(x_i, y_i)$ is the location of the $i^{th}$ region

and $n = \frac{\text{Area of the network}}{d}$

As per general means inequality, for n positive numbers $x_1, x_2 \ldots x_n$,

$$\frac{\sum_{i=1}^{n} x_i^2}{\sum_{i=1}^{n} x_i} \geq \left( \frac{1}{n} \sum_{i=1}^{n} x_i^2 \right)^{1/2} \geq \frac{1}{n} \sum_{i=1}^{n} x_i \quad \ldots (10)$$

Where the equality holds for $x_1 = x_2 = x_3 = \ldots = x_n$.

From equations (9) and (10), we obtain

$$\theta \geq \frac{1}{4} + \frac{\pi R^2}{4} \lim_{d \to 0} \left[ \frac{1}{n} \sum_{i=1}^{n} \rho(x_i, y_i) \right] = \frac{1}{4} + \frac{\pi R^2}{4} \lim_{d \to 0} \frac{\sum_{i=1}^{n} \rho(x_i, y_i) d}{nd}$$

$$= \frac{1}{4} + \frac{\pi R^2}{4} \frac{n}{\text{Area of the network}} \quad \ldots\ldots (11)$$

Where the equality holds for a uniform density $\rho(x,y)$ for all locations (x,y) in the network. Thus, the theoretical lower bound of the overhead ratio of MBMP-multi-hop to MBMP-power is $\frac{1}{4} + \frac{\pi R^2 n}{4 \times Area\ of\ the\ network}$, which can be achieved when density of nodes $\rho(x,y)$ is constant across the entire network. Hence, the higher the density of the network, the lower the overhead of MBMP-power approach as compared to that in MBMP-multi-hop approach. When the number of neighbors in a node's transmission range, $\pi R^2 \rho(x,y)$, exceeds 3, MBMP-multi-hop approach imposes a higher overhead than the MBMP-power approach.

The above analytical evaluation is verified through simulations under different node densities. In our simulations, we have used 10 to 70 mobile hosts, randomly distributed in an area of 1000 m x 1000 m. Five different scenarios are used for each density. The radio transmission range is 250m and the carrier-sensing range is 550m. The bandwidth of the channel is 2 Mbps. Ten randomly chosen pairs of nodes are used to establish a connection for a CBR traffic with transmission rate of 10 packets per second, the size of each packet being 512B. The mobility of nodes supports the random way-point model. The speed of nodes is 5 m/sec and the pause time is 20 sec. The simulations did run for 200 sec.

The overhead ratio of MBMP-multi-hop versus MBMP-power, obtained from theoretical analysis as well as the simulations, is depicted in Fig. 16. It can be observed that overhead ratio as per the simulations, is higher than the theoretical lower bound, as shown in Fig. 16. As per equation (11), when the nodes in the network are not evenly distributed, the overhead ratio becomes higher than the theoretical lower bound. We can also conclude from the simulations that MBMP-power has a comparatively less overhead than MBMP-multi-hop approach, when the average density of nodes in the ad hoc network is more than 15.3 nodes/$10^6 m^2$, which is too low a density to maintain connectivity of an ad hoc network as shown in [28].

## 7. EVALUATION:

Performance of MBMP is evaluated by us in this section through simulations using network simulator NS2 [20]. In the simulations, IEEE 802.11 [1] is used as the MAC layer protocol, as it does not support the QoS requirements. Hence, the results of simulations are oriented towards the effectiveness of MBMP, rather than that of QoS scheduling algorithms. MBMP is capable of providing guarantees in terms of average performance of flows over a short period of time (1 second) instead of instantaneous throughout / delay requirements of the flow, as IEEE 802.11 The QoS violation rates of MBMP-multi-hop, MBMP-power, MBMP-CS, SWAN and DSR are depicted in Fig. 27. It can be clearly observed that the QoS violations of all three versions of MBMP are very close to zero and overlap each other irrespective of the density of the network. But QoS violations in SWAN and DSR are observed to be much larger as compared to all three versions of MBMP (MBMP-multi-hop, MBMP-power, MBMP-CS) with DSR [5] and SWAN [4] for accuracy in bandwidth management and the incurred overhead.

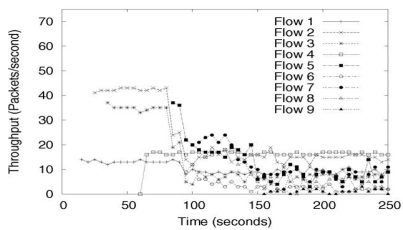   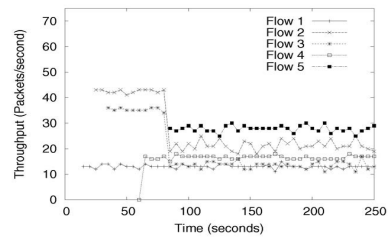

Fig.17: Throughput of DSR                              Fig.18: Throughput of SWAN

### 7.1. Effectiveness of MBMP:

In our simulations, we have used a 1000m x 1000m static network with 20 randomly positioned hosts to illustrate the effectiveness of MBMP. Nine CBR flows are attempted to establish in the network with randomly chosen source and destination nodes. Table-1 demonstrates the transmission rate, packet size and starting time of each of the nine scheduled CBR flows. The throughput and delay of the nine above flows with the implementation of DSR are depicted in

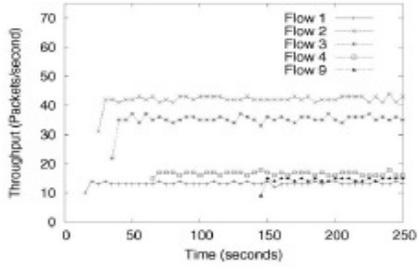

Fig.19: Throughput of MBMP-Multi-hop

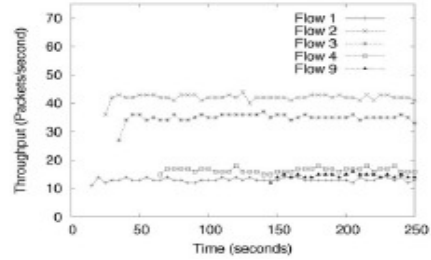

Fig.20: Throughput of MBMP-Power

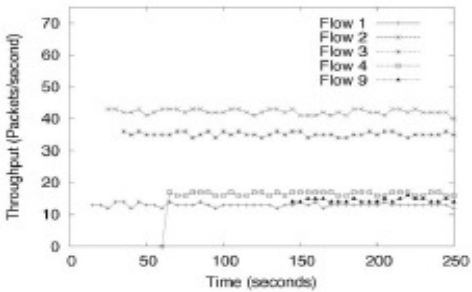

Fig.21: Throughput of MBMP-CS

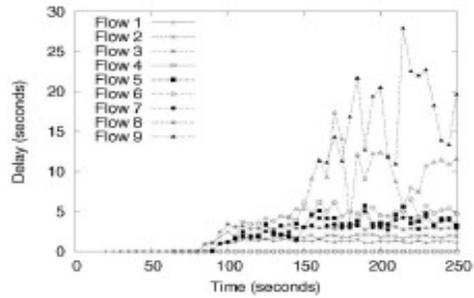

Fig.22: Delay of DSR

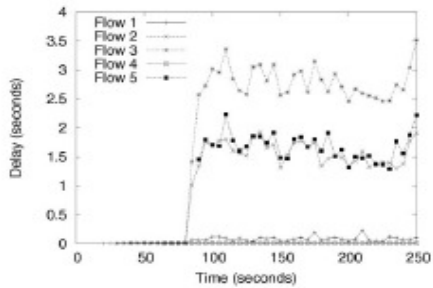

Fig.23: Delay of SWAN

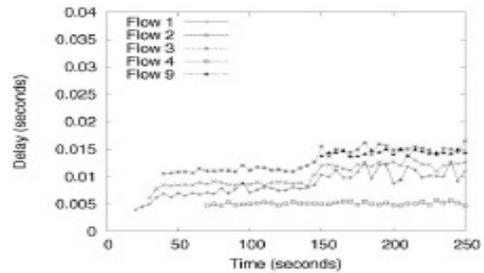

Fig.24: Delay of MBMP-multi-hop

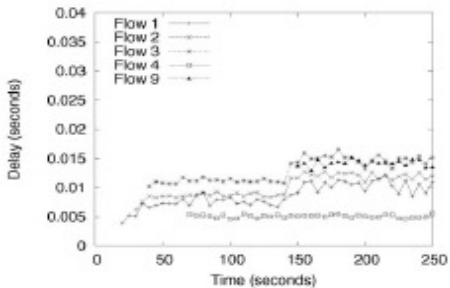

Fig.25: Delay of MBMP-Power

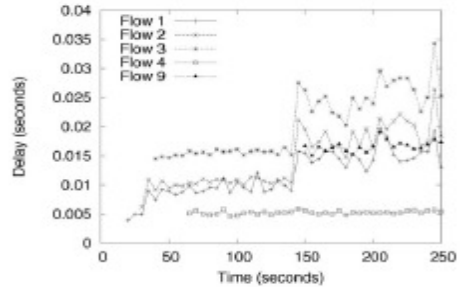

Fig.26: Delay of MBMP-CS

Fig. 17 and Fig. 22 respectively. As DSR does not implement any admission control procedure, addition of new flows results in congestion in the network, which consequently decreases throughput and dramatically increases the delay of the flows. Accordingly, Fig. 18 and Fig. 23 demonstrate the throughput and delay using SWAN respectively. It can be observed that only flows 1 to 5 are admitted by SWAN and the flows possess a more stable throughput and the delays of the flows are much lower as compared to that using DSR. But, with increased number of flows in the network, a significant degradation in throughput can be noticed and the delay increases significantly too, since SWAN does not take into account contentions between flows located in each other's c-neighborhood, which may lead to falsely admitted flows, and which may consequently affect QoS of existing flows, as depicted in the throughput and delay of flows 2 and 3 in Fig. 18 and Fig. 23 respectively. The throughput and delay of the flows in MBMP-multi-hop, MBMP-power and MBMP-CS are shown in figures [19, 20, 21, 24, 25, 26]. It can be observed that all three versions of MBMP maintain the throughput of the admitted flows. The worst-case delay of flows in all three versions of MBMP taken together is recorded to be below 35 ms, which is 100 times smaller than that of SWAN (i.e. 3.5 seconds), and 823 times smaller than the worst case delay of DSR (i.e. 28 seconds). It should be noted that the scales used in Fig. 22 and Fig. 23 are different from each other and are much larger than the scale used in figures 24,25,26.

## 7.2. Evaluation of Accuracy of Bandwidth Management:

The accuracy of admission control can be measured using two metrics: 1) number of false admissions; 2) bandwidth utilization. A false admission occurs when a flow is admitted, whose bandwidth consumption is beyond the available capacity of the network. The principal characteristic of a falsely admitted flow is that it either degrades the QoS requirements of already admitted flows or it is unable to achieve its own QoS requirements. Hence the number of false admissions can be assumed from the rate of QoS violations of admitted flows, which is the summation of actual throughput of admitted flows subtracting the summation of traffic generation rate of their CBR sources,

i.e. $N_f = \sum_{i=1}^{n} th_i - \sum_{i=1}^{n} tr_i$ ............... (12)

Where $N_f$ is the QoS violation parameter, $tr_i$ is the transmission rate of CBR source of $i^{th}$ flow, and $th_i$ is the actual throughput of $i^{th}$ flow. In fact, admission control should maintain QoS violation level to zero, and a negative value of QoS violation parameter indicates false admission. The second metric of evaluation of accuracy of admission control is bandwidth utilization. If an admission control can afford to reject a flow whose bandwidth consumption is very much within the capacity of the network, then the bandwidth of the network remains underutilized, by however, reducing the amount of traffic across the network. Thus, the total throughput of admitted flows in the network indicates the bandwidth utilization. We have carried out a comparison in our simulations, the total throughput and QoS violations of already admitted flows using all three versions of MBMP with that of the admission control algorithm used in SWAN [4]. The performance of DSR is also examined by us to demonstrate the necessity of admission control.

In our simulations, 450 1000 m x 1000 m networks are randomly generated. The number of nodes in the network range from 20 to 180. Each of the simulations did run for 200 seconds. Twenty randomly chosen pairs of nodes attempt to establish a connection with each other with a CBR traffic source. The transmission rates of CBR sources are uniformly distributed in [10, 50] packets per second with the packet size uniformly distributed in [100, 1000] bytes. For the mobility of the nodes, the random way point model is used, with maximum speed of 5 meters / second and a pause time of 10 seconds.

The QoS violation rates of MBMP-multi-hop, MBMP-power, MBMP-CS, SWAN and DSR are depicted in Fig. 27. It can be clearly observed that the QoS violations of all three versions of MBMP are very close to zero and overlap each other irrespective of the density of the network. But QoS violations in SWAN and DSR are observed to be much larger as compared to all three versions of MBMP, which indicates that, there could be more false admissions in SWAN and

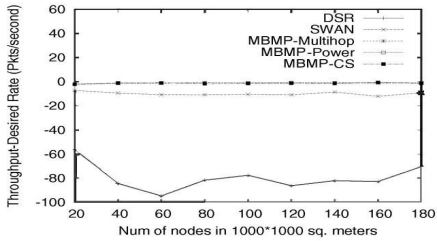 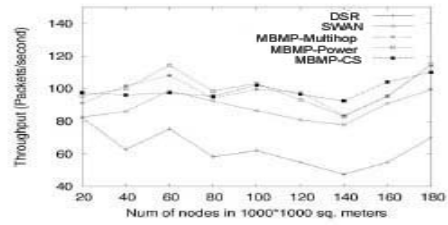

Fig.27: Rate of QoS violation                    Fig.28: Network total throughput

DSR. The total throughput of admitted flows in MBMP-multi-hop, MBMP-power, MBMP-CS, SWAN and DSR is shown in Fig. 28. It can be obviously observed that the throughput of all three versions of MBMP is much larger than that of SWAN and DSR. However, the throughput of SWAN is close to that of MBMP-CS only when the density of nodes in the network is 60 nodes per $10^6 m^2$, but it is still smaller than the throughput of MBMP-multi-hop and MBMP-power. It indicates that the bandwidth utilization in MBMP is high as the capacity of the network is not reduced in any case. In addition to this, all three versions of MBMP convincingly reduce the amount of collisions among admitted flows as it has fewer false admissions, which consequently increases the capacity of the network. It is also observed that throughput of MBMP-CS is lower than that of MBMP-multi-hop and MBMP-power when the density of nodes in the network is low, which is a consequence of the conservative c-neighborhood bandwidth estimation (section 4.1.2), used in MBMP-CS approach. But, with increasing density of nodes in the network, the message overhead involved in active c-neighborhood bandwidth estimation used in MBMP-multi-hop and MBMP-power approaches increases convincingly, which consumes more network capacity and causes a prominent reduction of throughput, and thus, throughput of MBMP-multi-hop and MBMP-power falls below the throughput of MBMP-CS, as bandwidth estimation in MBMP-CS involves negligibly small overhead even in higher densities of nodes in the network. Hence, it can be concluded that in dense networks, throughput of MBMP-CS exceeds that of MBMP-multi-hop and MBMP-power approaches.

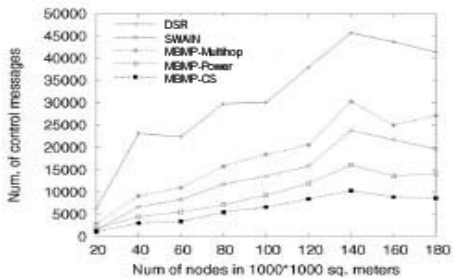 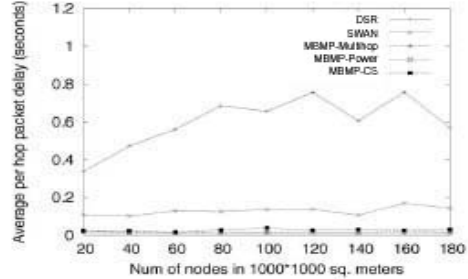

Fig.29: Control message overhead              Fig.30: Average per hop packet delay

### 7.3. Message Overhead Involved In Admission Control:

In active admission control methods, like MBMP-multi-hop and MBMP-power, the amount of control messages is increased as they require c-neighbors to exchange control messages for admission control during admission request and admission rejection phases. On the other hand, an effective admission control reduces the number of control messages used during route discovery for two reasons. 1) the admission control performed during route request phase can initially eliminate the routes with insufficient bandwidth for the flow, causing a reduction of route requests in the hot spots; 2) by preventing the network from being overloaded, an effective admission control can reduce the number of link failures caused due to collisions between neighboring nodes, which consequently reduces the number of control messages caused due to re-establishment of routes.

For evaluation of effectiveness of MBMP with respect to control message overhead, the number of control messages used in the simulations in section 7.2, including route request, route reply, admission request and admission rejection messages, is recorded and compared with DSR and SWAN. The number of control messages during the simulations, is depicted in Fig. 29. It can be noticed from Fig. 29, that the control message overhead of DSR is the largest, which demonstrates that the overall effect of admission control can reduce control message overhead. The control message overhead of MBMP-multi-hop is larger than SWAN due to its extra message overhead of admission request and admission rejection messages. As demonstrated in section 6, the control message overhead in MBMP-power is smaller than that of MBMP-multi-hop. MBMP-power possesses less message overhead as compared to that of SWAN too, since it reduces control message overhead for route discovery, thereby compensating its extra message overhead of communication with its c-neighbors. MBMP-CS possesses the lowest message overhead, since it does not impose any extra control messages except the messages used for route discovery. It could be observed from the simulations, that even though MBMP implements varieties of control messages to perform admission control, its bandwidth-aware routing reduces the total control message overhead. As depicted in Fig. 28, MBMP never reduces the capacity of the network, and hence, its message overhead is very much acceptable.

Although MBMP is purposefully designed for efficient bandwidth management, its bandwidth-aware routing strategy essentially balances the load in the network, as routes through hot spots are usefully rejected. Thus, even though MBMP provides higher throughput than SWAN and DSR, it has control over the delay suffered by the packets in the network, by avoiding congested areas in the network. The average per hop delay of a packet in the simulations (section 7.2) is demonstrated in Fig. 30. It can be observed from Fig. 30, that all three versions of MBMP achieve much lower packet delay than SWAN and DSR, which ultimately indicates the excellent capability of MBMP to balance the load in the network.

## 8. CONCLUSION AND FUTURE WORK:

In our current research paper, we have presented three methods to achieve bandwidth management in multi-hop mobile ad hoc networks with admission control. Major contribution of this paper is the inclusion of information from nodes within the carrier-sensing range and outside transmission range during the process of admission control. As demonstrated through simulations, MBMP effectively manages requests for bandwidth even beyond the capacity of the network, and consequently reduces the control message overhead on the network. However, it performs only the admission control part of QoS protocol stack. It can be combined with many existing QoS protocols, such as QoS-aware MAC protocols. Different QoS-aware MAC protocols and admission policies may affect the definition of local available bandwidth so that extensions of equations (1) & (2) may be required for estimation of local / c-neighborhood available bandwidth.

## REFERENCE:

[1] IEEE Computer Society, "802.11: Wireless LAN Medium Access Control (MAC) and Physical Layer (PHY) Specifications," 1999.
[2] S. Mangold, S. Choi, P. May, O. Klein, G. Hiertz, and L. Stibor,"IEEE 802.11e Wireless LAN for Quality of Service," Proc.European Wireless, 2002.
[3] R. Rozovsky and P.R. Kumar, "SEEDEX: A MAC Protocol for AdHoc Network," Proc. ACM Symp. Mobile Ad Hoc Networking &Computing, 2001.
[4] G.-S. Ahn, A. Campbell, A. Veres, and L.-H. Sun, "SWAN: ServiceDifferentiation in Stateless Wireless AdHoc Networks," Proc. Infocom, 2002.
[5] D.B. Johnson and D.A. Maltz, "Dynamic Source Routing in Ad Hoc Wireless Networks," Mobile Computing, vol. 353, 1996.
[6] D. Maltz, "Resource Management in Multi-Hop Ad Hoc Networks," Technical Report CMU CS 00-150, School of Computer Science, Carnegie Mellon Univ., July 2000.

[7] Y.-C. Hsu and T.-C. Tsai, "Bandwidth Routing in Multihop Packet Radio Environment," Proc. Third Int'l Mobile Computing Workshop,1997.
[8] T.-W. Chen, J.T. Tsai, and M. Gerla, "QoS Routing Performance in Multihop Multimedia Wireless Networks," Proc. IEEE Int'l Conf. Universal Personal Comm. (ICUPC), 1997. 376 IEEE TRANSACTIONS ON MOBILE COMPUTING, VOL. 4, NO. 4, JULY/AUGUST 2005
[9] C.R. Lin and C.-C. Liu, "An On-Demand QoS Routing Protocol for Mobile Ad Hoc Networks," IEEE Global Telecomm. Conf., 2000.
[10] C. Zhu and M.S. Corson, "QoS Routing for Mobile Ad Hoc Networks," Technical Report CSHCN TR 2001-18, Inst. for System Research, Univ. of Maryland, 2001.
[11] C.R. Lin and J.-S. Liu, "QoS Routing in Ad Hoc Wireless Networks," IEEE J. Selected Areas in Comm., vol. 17, no. 8, pp. 1426-1438, Nov. /Dec. 1999.
[12] V. Kanodia, C. Li, A. Sabharwal, B. Sadeghi, and E. Knightly, "Distributed Multi-Hop Scheduling and Medium Access with Delay and Throughput Constraints," Proc. Seventh Ann. Int'l Conf. Mobile Computing and Networking, 2001.
[13] H. Luo, S. Lu, V. Bharghavan, J. Cheng, and G. Zhong, "A Packet Scheduling Approach to QoS Support in Multihop Wireless Networks," ACM J. Mobile Networks and Applications, special issue on QoS in heterogeneous wireless networks, 2002.
[14] S.-B. Lee, G.-S. Ahn, X. Zhang, and A. Campbell, "INSIGNIA: An IP-Based Quality of Service Framework for Mobile Ad Hoc Networks," J. Parallel and Distributed Computing(PDC), special issue on wireless and mobile computing and communications, vol. 60, pp. 374-406, 2000.
[15] M.G. Barry, A.T. Campbell, and A. Veres, "Distributed Control Algorithms for Service Differentiation in Wireless Packet Networks," Proc. Infocom, 2001.
[16] R. Ramanathan and M. Steenstrup, "Hierarchically-Organized, Multihop Mobile Wireless Networks for Quality-of-Service Support," Mobile Networks and Applications (MNA), vol. 3, no.1, pp.101-119,1998.
[17] S. Murthy and J.J. Garcia-Luna-Aceves, "A Routing Architecture for Mobile Integrated Services Networks," Mobile Networks and Applications (MNA), vol. 3, no. 4, pp. 391-407, 1999.
[18] S.H. Shah, K. Chen, and K. Nahrstedt, "Dynamic Bandwidth Management for Single-Hop Ad Hoc Wireless Networks," Proc. IEEE Int'l Conf. Pervasive Computing and Comm., 2003.
[19] M. Kazantzidis, M. Gerla, and S.-J. Lee, "Permissible Throughput Network Feedback for Adaptive Multimedia in AODV MANETs," IEEE Int'l Conf. Comm., 2001.
[20] K. Fall and K. Varadhan, "NS Notes and Documentation," The VINT Project, UC Berkeley, LBL, USC/ISI, and Xerox PARC, 1997.
[21] Binod Kumar Patanayak, Alok Kumar Jagadev, Manoj Kumar Mishra, Dr. Manojranjan Nayak, "Evaluation of A Centralized Bandwidth Management Protocol Architecture with Admission Control In Mobile Ad Hoc Networks", International Journal on Computer Engineering and Information Technology(IJCEIT), Volume 2, No. 2, PP 79-91,Nov 2008-Jan 2009.
[22] G. Bianchi, "Performance Analysis of the IEEE 802. 11 Distributed Coordination Function," IEEE J. Selected Areas in Comm.(SAC), vol. 18, no. 3, 2000.
[23] Y. Yang, J. Wang, and R. Kravets, "Achievable Bandwidth Prediction in Multihop Wireless Networks," Technical Report UIUCDCS-R-2003-2367, Dec. 2003.
[24] P. Gupta and P.R. Kumar, "Capacity of Wireless Networks," IEEE Trans. Information Theory(IT), no. 2, pp. 388-404, 2000.
[25] C. Perkins and P. Bhagwat, "Highly Dynamic Destination- Sequenced Distance-Vector Routing (DSDV) for Mobile Computers," ACM SIGCOMM '94 Conf. Comm. Architectures, Protocols and Applications, 1994.
[26] C. Perkins, "Ad-Hoc On-Demand Distance Vector Routing," Proc. Military Comm. Conf., 1997.
[27] V.D. Park and M.S. Corson, "A Highly Adaptive Distributed Routing Algorithm for Mobile Wireless Networks," Proc. INFOCOM, pp. 1405-1413, 1997.
[28] O. Dousse, P. Thiran, and M. Hasler, "Connectivity in Ad-Hoc and Hybrid Networks," Proc. INFOCOM, pp. 1079-1088, June 2002.
[29] Y. Hwang and P. Varshney, "An Adaptive QoS Routing Protocol with Dispersity for Ad-hoc Networks," Proc.. 36[th] Hawali Int'lConf. Sys. Sci., Jan. 2003.
[30] Y. Ge, T. Kunz and L. Lamont, "Quality of Service Routing in Ad-hoc Networks using OLSR", Proc.. 36[th] Hawali Int'lConf. Sys. Sci., Jan. 2003.